\documentstyle[fleqn]{article}
\begin{document}


 \title{Staggered Polarization of Vertex Models \\
          with $U_q(\widehat{sl}(n))$-Symmetry }
 \author{Yoshitaka Koyama \\ \\
 Research Institute for Mathematical sciences,  \\
 Kyoto University, Kyoto 606, Japan}
 \date{}
 \maketitle

 \begin{abstract}

  In this paper we give
  an explicit formula for
  level 1 vertex operators related to
  $U_q(\widehat{sl}(n))$
  as operators on the Fock spaces.
  We derive also
  their commutation relations.
  As an applications
  we culculate
  the one point functions of
  the one-dimensional spin chain
  associated with the vector representation of
  $U_q(\widehat{sl}(n))$,
  thereby extending
  the recent work on
  the staggered polarization of the $XXZ$-model.

 \end{abstract}


  \section{Introduction}

   The Hamiltonian of the $XXZ$-model has
   $U_q(\widehat{sl}(2))$-symmetry
   in the thermodynamic limit.
   Recently,
   on the basis of this fact,
   the $XXZ$-model was formulated
   in the framework of
   representation theory of $U_q(\widehat{sl}(2))$.
   Let us explain the scheme described in \cite{dfjmn} briefly.

   First we recall the $XXZ$-model as it appears in physics.
   The space of states of the $XXZ$-model is
   the infinite tensor product
   $ \cdots \otimes V \otimes V \otimes V \otimes \cdots $,
   where
   $ V={\bf C}v_+ \oplus {\bf C} v_- $
   is the two-dimensional vector space.
   The $XXZ$-Hamiltonian is the following operator
   formally acting the above space:
   $$ H_{XXZ}=-\frac{1}{2}\sum_{k \in {\bf Z}}
                 (\sigma^x_k \sigma^x_{k+1}
                 +\sigma^y_k \sigma^y_{k+1}
                 +\frac{q+q^{-1}}{2}\sigma^z_k \sigma^z_{k+1}) , $$
   where
   $ \sigma^x,\sigma^y,\sigma^z $
   are the Pauli matrices on V,
   $\sigma_k^{\alpha}$ acting on the k-th component of
   $ \cdots \otimes V \otimes V \otimes V \otimes \cdots $.
   Let
   $U'_q(\widehat{sl}(2))$
   denote the subalgebra of
   $U_q(\widehat{sl}(2))$
   with the grading operator $d$ being dropped.
   It acts on $V$ as follows.
   $$ e_1.v_-=v_+ , \;
      f_1.v_+=v_- , \;
      t_1.v_{\pm}=q^{\pm 1} v_{\pm} , $$
   $$ e_0.v_+=v_- , \;
      f_0.v_-=v_+ , \;
      t_0.v_{\pm}=q^{\mp 1} v_{\pm} . $$
   Furthermore,
   $U'_q(\widehat{sl}(2))$
   acts on
   $ \cdots \otimes V \otimes V \otimes V \otimes \cdots $
   via the iterated coproduct
   $ \Delta^{(\infty)} $.
   $$ \Delta^{(\infty)} (t_i)
             = \cdots
               \otimes t_i \otimes t_i \otimes t_i \otimes
               \cdots , $$
   $$ \Delta^{(\infty)} (e_i)
             = \sum
               \cdots
               \otimes t_i \otimes t_i \otimes e_i
               \otimes 1 \otimes 1 \otimes
               \cdots , $$
   $$ \Delta^{(\infty)} (f_i)
            =  \sum
               \cdots
               \otimes 1 \otimes 1 \otimes f_i
               \otimes t_i^{-1} \otimes t_i^{-1} \otimes
               \cdots , $$
   Formal manipulation shows that
   $$ [ \, H_{XXZ} \, , \, U'_q(\widehat{sl}(2)) \, ]=0 . $$
   Letting $T$ denoteb the shift operator on
   $ \cdots \otimes V \otimes V \otimes V \otimes \cdots $,
   we can also check
   $$ \frac{2q}{1-q^2} H_{XXZ}=T^2 d T^{-2} - d . $$

   The above observation holds
   only in the infinite lattice case.
   Of course,
   $H_{XXZ}$ and the action of $U_q(\widehat{sl}(2))$
   are not literally well-defined.
   Nevertheless,
   when we consider the model
   in the anti-ferroelctric regime $-1<q<0$,
   we can construct a well-defined theory
   on ``the space of physical states'',
   which is the subspace consisting of
   finite excitations over the ground states in
   $ \cdots \otimes V \otimes V \otimes V \otimes \cdots $.
   The formulation of \cite{dfjmn}
   is based on the (hypothetical) identification
   $$ {\bf ``the \; space \; of \; physical \; states''}
      = \bigoplus_{0 \leq i,j \leq 1}
           V(\Lambda_i) \otimes V(\Lambda_j)^* , $$
   where $V(\Lambda_i)$ is
   the level 1 highest weight irreducible
   $U_q(\widehat{sl}(2))$-module
   and
   $V(\Lambda_j)^*$
   is the dual module of
   $V(\Lambda_j)$.
   The symbol $\otimes$
   is to be understood
   with an appropriate completion,
   but we will not go into such details in the sequel.
   To motivate this hypothesis,
   consider the intertwiner of
   $U_q(\widehat{sl}(n))$-modules
   $$ \tilde{\Phi}^{\Lambda_{1-i} V}_{\Lambda_i}
      \; : \;
      V(\Lambda_i)
         \longrightarrow V(\Lambda_{1-i}) \otimes V ,
      \qquad (*) $$
   called vertex operators(\cite{fr}).
   In fact,
   such an operator exists,
   is unique up to a scaler,
   gives an isomorphosm.
   Iterating the vertex operators,
   we get the following isomorphism.
   $$ V(\Lambda_i) \otimes V(\Lambda_j)^*
      \cong
      V(\Lambda_{1-i}) \otimes V \otimes V(\Lambda_j)^*
      \cong
      V(\Lambda_{0 {\rm or} 1})
      \otimes V \otimes
      \cdots \otimes V \otimes V(\Lambda_j)^* . $$
   It tells us
   that the local structure
   $ \cdots \otimes V \otimes V \otimes \cdots $
   in the naive picture
   is realized in the space
   $ \bigoplus V(\Lambda_i) \otimes V(\Lambda_j)^* $.
   By composing $(*)$ with a similar vertex operator
   $$ V \otimes V(\Lambda_i)^*
         \longrightarrow V(\Lambda_{1-i})^* , $$
   we get
   $$ V(\Lambda_i) \otimes V(\Lambda_j)^*
      \cong
      V(\Lambda_{1-i}) \otimes V \otimes V(\Lambda_j)^*
      \cong
      V(\Lambda_{1-i}) \otimes V(\Lambda_{1-j})^* . $$
   The resulting isomorphism
   can be identified with
   the shift operator $T$.
   In this manner
   we can build a well-defined theory on
   $ \bigoplus V(\Lambda_i) \otimes V(\Lambda_j)^* $
   that captures all the essential features
   expected from the physical definition.

   It is straightforward
   to generalize the above formulation
   to the models related to
   any quantum affine algebra.
   In this paper,
   we consider
   a multi spin-analogue of the $XXZ$-model
   related to
   the vector representation of
   $U_q(\widehat{sl}(n))$.
   Our main results are twofold.
   One is
   the bosonization of
   the level 1 vertex operators({\bf Theorem 3.3, 3.4}).
   The other is
   the exact calculation of the one-point functions({\bf Theorem 5.2}).
   The structure of this paper is the following.
   In \S 2,
   we review the construction of
   the level 1 irreducible highest weight
   $U_q(\widehat{sl}(n))$-modules.
   In \S 3,
   we construct
   the vertex operators
   on the bosonic Fock space explicitly.
   In \S 4,
   we explain
   the mathematical formulation of models.
   In \S 5,
   first, we derive
   an integral representation
   for the one-point function
   by using
   the bosonization of the vertex operators.
   Next,
   by using
   the commutation relations
   of the vertex operators,
   we derive
   difference equations
   for the one-point functions.
   This equation can be solved easily.
   As a result,
   we obtain
   an explicit formula of the one-point functions
   exteding th previous work on
   the spontaneous staggered polarization
   for the $XXZ$-model(\cite{b}).


  \section{Vertex operator representations of $U_q(\widehat{sl}(n))$}

   In this section,
   we review
   the construction of
   the level 1 irreducible highest weight modules
   following \cite{fj}.


   \subsection{Notations}

   Throughout this paper, we fix a real number q ($-1<q<0$)
   and a positive integer $n$.
   The $q$-integer and $q$-factorial
   are denoted by
   $ [k]=(q^k-q^{-k})/(q-q^{-1}) $
   and
   $ (a;q)_{\infty}=\prod^{\infty}_{k=0}(1-aq^k) $
   respectively.
   Most notations concerning Lie algebras follow \cite{k}.
   Let $P$ be a free {\bf Z}-module
   $$ P:=\bigoplus^{n-1}_{i=0}
          {\bf Z} \Lambda_i \oplus {\bf Z} \delta . $$
   We call it the weight lattice.
   We define $P^*$ as follows.
   $$ P^*:={\rm Hom}(P,{\bf Z})
          =\bigoplus^{n-1}_{i=0}
                     {\bf Z} h_i \oplus {\bf Z} d . $$
   The pairing is given by
   $ \langle \Lambda_i , h_j \rangle =\delta_{ij} $,
   $ \langle \Lambda_i , d \rangle =0             $,
   $ \langle \delta , h_j \rangle =0              $,
   $ \langle \delta , d \rangle =1                $.
   The indices are extended cyclically such as
   $ \Lambda_i=\Lambda_{i+n} $, etc.
   Let
   $ \alpha_0
        = -\Lambda_{n-1} +2 \Lambda_0 -\Lambda_1 +\delta $,
   $ \alpha_j
        = -\Lambda_{j-1}
            +2 \Lambda_j -\Lambda_{j+1} $
   $ (1 \leq j \leq n-1 ) $
   be the simple roots.
   The invariant bilinear form on $P$
   is given by
   $ (\alpha_i|\alpha_j)
                 =-\delta_{ij-1}+2\delta_{ij}-\delta_{ij+1} $
   and
   $ ( \delta | \delta )=0 $.
   The projection to the classical weight lattice
   is given by
   $ \overline{\Lambda}_i=\Lambda_i-\Lambda_0 $,
   $ \overline{\delta}=0 $.
   $ U_q(\widehat{sl}(n)) $
   is the {\bf C}-algebra generated
   by the symbols
   $ \{ t_i^{\pm}(=q^{\pm h_i}),
        q^d, e_i, f_i, (i=0,\cdots,n-1) \} $
   which satisfy the following defining relations.
   $$ t_i t_j = t_j t_i
      \; , \;
      t_i e_j t_i^{-1}=q^{\langle \alpha_j,h_i \rangle} e_j
      \; , \;
      t_i f_j t_i^{-1}=q^{- \langle \alpha_j,h_i \rangle} f_j $$
   $$ [e_i,f_j] = \delta_{ij} \frac{t_i-t_i^{-1}}{q-q^{-1}} $$
   $$ \sum_{k=0}^b (-1)^k
          \left[
             \begin{array}{l} b \\ k \end{array}
          \right]
         e_i^k e_j e_i^{b-k}=0
      \; , \;
      \sum_{k=0}^b (-1)^k
          \left[
             \begin{array}{l} b \\ k \end{array}
          \right]
         f_i^k f_j f_i^{b-k}=0 $$
   where
   $ b=1- \langle \alpha_i,h_j \rangle $,
   $      \left[
             \begin{array}{l} b \\ k \end{array}
          \right]
          =\displaystyle{\frac{[b]!}{[k]![b-k]!}} $,
   $ [k]!=[1][2] \cdots [k] $.
   Throughout this paper,
   we denote
   $ U_q(\widehat{sl}(n)) $
   by $U_q$.
   $ U'_q $
   is the subalgebra of $U_q$ generated by
   $ \{ t_i,e_i,f_i \} $.
   We denote
   the irreducible highest weight
   $U_q$(or $U'_q$)-module with highest weight $\lambda$
   by $V(\lambda)$.
   We fix a highest weight vector of $V(\lambda)$
   and denote it by $|\lambda \rangle$.
   The coproduct $\Delta$ and antipode $S$
   are given as follows.
   $$ \Delta (q^h) = q^h \otimes q^h
      \; , \;
      \Delta (e_i) = e_i \otimes 1 + t_i \otimes e_i
      \; , \;
      \Delta (f_i) = f_i \otimes t_i^{-1} + 1 \otimes f_i \; , $$
   $$ S(q^h)=q^{-h}
      \; , \;
      S(e_i) = -t_i^{-1}e_i
      \; , \;
      S(f_i)= -f_i t_i \; . $$
   When $W$ is a $U_q$(or $U'_q$)-module,
   we introduce the left module structure
   on the dual space $W^*$
   by
   $x.u^*(v)=u^*(S(x).v)$ for $x \in U_q$(or $U'_q$),
   $u^* \in W^*$ and $v \in W$.
   If $W$ has a weight decomposition
   $ \bigoplus_{\lambda} W_{\lambda} $,
   we define the completion
   $ \widehat{W} = \prod_{\lambda} W_{\lambda} $.
   Normally we omit $ \; \widehat{\;} \; $.


   \subsection{Drinfeld generators of $U_q$}

    We introduce another set of generators of $U_q$ (\cite{d}).
    \\
    {\bf Definition.}
    {\it
     ${\cal A}$ is the {\bf C}-algebra
     generated by the symbols
     $ \{ \, \gamma^{\pm \frac{1}{2}}, \, K_i, \, a_i(k), \, x^{\pm}_i(l)
             \\ ( 1 \leq i \leq n-1, \,
           k \in {\bf Z} \setminus \{ 0 \}, \, l \in {\bf Z} ) \} $
     which satisfy the following defining relations.}
     \[ 1) \hspace{.75in}
            \gamma^{\pm\frac{1}{2}}
                   \in {\it Center \; of} \; {\cal A} \; , \quad
            \gamma^{\frac{1}{2}} \gamma^{-\frac{1}{2}}=1 \; ,              \]
     \[ 2) \hspace{.75in}
            [a_i(k),a_j(l)]
              =\delta_{k+l,0} \frac{[(\alpha_i | \alpha_j)k]}{k}
                              \frac{\gamma^k - \gamma^{-k}}{q-q^{-1}} \; , \]
     \[ 3) \hspace{1.5in}
            [a_i(k),K_j]=0 \; , \]
     \[ 4) \hspace{1.25in}
            K_i x^{\pm}_j(k) K_i^{-1}
                =q^{\pm \langle \alpha_j , h_i \rangle} x^{\pm}_j(k) \; ,  \]
     \[ 5) \hspace{1in}
            [a_i(k),x^{\pm}_j(l)]
                =\pm \frac{[\langle \alpha_j ,h_i \rangle k]}{k}
                     \gamma^{\mp \frac{|k|}{2}} x^{\pm}_j(k+l) \; , \]
     \[ 6) \hspace{.5in} x^{\pm}_i(k+1) x^{\pm}_j(l)
                        -q^{\pm (\alpha_i | \alpha_j)}
                                  x^{\pm}_j(l) x^{\pm}_i(k+1)       \]
     \[    \hspace{1.5in}=q^{\pm (\alpha_i | \alpha_j)}
                                 x^{\pm}_i(k) x^{\pm}_j(l+1)
                                -x^{\pm}_j(l+1) x^{\pm}_i(k)   \; , \]
     \[ 7) \hspace{.4in} [x^+_i(k),x^-_j(l)]
                =\frac{\delta_{ij}}{q-q^{-1}}
                 (\gamma^{(k-l)/2} \psi_i(k+l)
                             -\gamma^{(l-k)/2} \varphi_i(k+l))      \]
           \hspace{.7in} {\it where}
     \[    \hspace{.5in}
                 \left \{ \begin{array}{l}
                 \displaystyle{
                      \sum^{\infty}_{k=0} \psi_i(k) z^{-k}
                     =K_i \, {\rm exp} \left( (q-q^{-1})
                             \sum^{\infty}_{k=1} a_i(k) z^{-k}
                                                       \right) \; } ,
                 \\
                 \displaystyle{
                      \sum^{\infty}_{k=0} \varphi_i(-k) z^{-k}
                     =K_i^{-1} {\rm exp} \left( -(q-q^{-1})
                                  \sum^{\infty}_{k=1} a_i(-k) z^{-k}
                                                       \right) \; . }
                 \end{array} \right.   \]
     \[ 8) \hspace{.75in}
                  [x^{\pm}_i(k),x^{\pm}_j(l)]=0
                                   \; \; \; \; {\it for} \;
           \langle \alpha_i , h_j \rangle =0 \; ,   \]
     \[ 9) \]
     \[ \;     \{   x^{\pm}_i(k) x^{\pm}_i(l) x^{\pm}_j(m)
              -(q+q^{-1}) x^{\pm}_i(k) x^{\pm}_j(m) x^{\pm}_i(l)
                       + x^{\pm}_j(m) x^{\pm}_i(k) x^{\pm}_i(l) \} \]
     \[ \;    +\{   x^{\pm}_i(l) x^{\pm}_i(k) x^{\pm}_j(m)
              -(q+q^{-1}) x^{\pm}_i(l) x^{\pm}_j(m) x^{\pm}_i(k)
                       + x^{\pm}_j(m) x^{\pm}_i(l) x^{\pm}_i(k) \} \]
     \[ \: =0 \qquad \qquad {\it for} \;
                           \langle \alpha_j , h_i \rangle =-1 \; . \]
     \hspace{\fill} $\Box$
    \\
    We know the following theorem.
    \\
    {\bf Theorem 2.2}(\cite{d})
    {\it The following correspondance gives an isomorphism}
    $ U'_q \cong {\cal A} $.
    $$ t_j \longmapsto K_j      \; \; , \; \;
       e_j \longmapsto x^+_j(0) \; \; , \; \;
       f_j \longmapsto x^-_j(0)
           \hspace{.35in} (1 \leq j \leq n)    \; , $$
    $$ t_0 \longmapsto
           \gamma K_1^{-1} \cdots K_{n-1}^{-1} \; , $$
    $$ e_0 \longmapsto
           [x^-_{n-1}(0),[x^-_{n-2}(0),\cdots [x^-_2(0),x^-_1(1)]
           _{q^{-1}} \cdots ]_{q^{-1}}]_{q^{-1}}
           K_1^{-1} \cdots K_{n-1}^{-1}        \; , $$
    $$ f_0 \longmapsto
           K_1 \cdots K_{n-1}
           [[ \cdots [x^+_1(-1),x^+_2(0)]_q
              \cdots ,x^+_{n-2}(0)]_q,x^+_{n-1}(0)]_q  \; . $$

    {\it Here we have set} $ [A,B]_q=AB-qBA $.
    \hspace{\fill} $\Box$


   \subsection{Group algebra {\bf C}$[\overline{P}]$}

    For the construction of representations,
    it is enough
    to consider only {\bf C}$[\overline{Q}]$,
    where $\overline{Q}=\oplus^{j=n-1}_{j=1} \alpha_j$
    is the classical root lattice.
    But, for the construction of the vertex operators
    it is convinient to define {\bf C}$[\overline{P}]$
    ($\overline{P}=
           \oplus^{j=n-1}_{j=2} \alpha_j
           \oplus \overline{\Lambda}_{n-1}$
     : the classical weight lattice).
    In fact, we use
    a central-extention of the group algebra of
    $ \overline{P} $.
    \\
    {\bf Definition.}
    {\it
     {\bf C}$[\overline{P}]$ is
     the {\bf C}-algebra
     generated by the symbols
     \{ $e^{\alpha_2}, \, \cdots, \, e^{\alpha_{n-1}},
                      \\ e^{\overline{\Lambda}_{n-1}}$ \}
     which satisfy the following defining relations.}
     $$ e^{\alpha_i}e^{\alpha_j}
                 =(-1)^{(\alpha_i|\alpha_j)}
                     e^{\alpha_j}e^{\alpha_i}
                       \hspace{.75in} (2 \leq i \leq n-1)       $$
     $$ e^{\alpha_i}e^{\overline{\Lambda}_{n-1}}
                 =(-1)^{\delta_{in-1}}
                     e^{\overline{\Lambda}_{n-1}}e^{\alpha_i}
                       \hspace{.34in} (2 \leq i \leq n-1)       $$
     \hspace{\fill} $\Box$

    For $\alpha=m_2 \alpha_2 + \cdots + m_{n-1} \alpha_{n-1}
            +m_n \overline{\Lambda}_{n-1} (\in \overline{P})$,
    we denote
    $ e^{m_2 \alpha_2} \cdot \cdot \cdot e^{m_{n-1} \alpha_{n-1}}
                                        e^{\overline{\Lambda}_{n-1}} $
    by $ e^{\alpha} $.
    For example,
    $ e^{\alpha_1}
      =e^{-2 \alpha_2} e^{-3 \alpha_3}
       \cdots
       e^{-(n-1) \alpha_{n-1}} e^{n \overline{\Lambda}_{n-1}}   $,
    $ e^{\overline{\Lambda}_i}
      =e^{- \alpha_{i+1}} e^{-2 \alpha_{i+2}}
       \cdots
       e^{-(n-i-1) \alpha_{n-1}} e^{n \overline{\Lambda}_{n-1}} $.
    A simple culculation shows the following.
    \\
    {\bf Proposition 2.3}
    \[ 1) \hspace{.5in}
             e^{\alpha_i}e^{\alpha_j}
            =(-1)^{(\alpha_i|\alpha_j)}e^{\alpha_j}e^{\alpha_i}
          \quad (1 \leq i,j \leq n-1) \]
    \[ 2) \hspace{.5in}
             e^{\alpha_1}e^{\overline{\Lambda}_{n-1}}
            =(-1)^{n-1}e^{\overline{\Lambda}_{n-1}}e^{\alpha_1}           \]
    \[ 3) \hspace{.5in}
             e^{\alpha_i}e^{\overline{\Lambda}_1}
            =(-1)^{n \delta_{i1}}e^{\overline{\Lambda}_1}e^{\alpha_i}
          \quad (1 \leq i \leq n-1)   \]
    \[ 4) \hspace{.5in}
             e^{\overline{\Lambda}_1}e^{\overline{\Lambda}_{n-1}}
            =(-1)^n e^{\overline{\Lambda}_{n-1}}e^{\overline{\Lambda}_1}  \]
    \hspace{\fill} $\Box$


   \subsection{Construction of representations}
   Let
   $$ W_i:={\bf C}[a_j(-k) (1 \leq j \leq n-1,k \in {\bf Z}_{>0})]
                 \otimes {\bf C}[\overline{Q}]e^{\overline{\Lambda}_i}
                                       \hspace{.2in} (0 \leq i \leq n-1). $$
   We define the operators
   $ a_j(k) \;  ( 1 \leq j \leq n-1 ) $ ,
   $ \partial_\alpha $ ,
   $ e^\alpha (\alpha \in \overline{Q}) $ ,
   $ d $
   on $W_i$ as follows:
   \\
   for
   $ f \otimes e^\beta
        =a_{i_1}(-n_1) \cdots a_{i_k}(-n_k) \otimes e^\beta \in W_i $,
   $$ a_j(k).f \otimes e^\beta
                = \left \{ \begin{array}{ll}
                               a_j(k)f \otimes e^\beta      & (k<0) \\
                      \mbox{$[a_j(k),f] \otimes e^\beta$}   & (k>0)
                           \end{array} \right.                        $$
   \[ \hspace{.75in}
      \partial_\alpha .f \otimes e^\beta
                =(\alpha|\beta)f \otimes e^\beta  \]
   \[ \hspace{.75in}
      e^\alpha .f \otimes e^\beta
                =f \otimes e^\alpha e^\beta       \]
   $$ d.f \otimes e^\beta
        =(-\sum^k_{l=1} n_l-\frac{(\beta|\beta)}{2}
                   +\frac{(\overline{\Lambda_i}|\overline{\Lambda_i})}{2})
         f \otimes e^\beta . $$
   \\
   Let
   $$ X^{\pm}_j(z)
       :=\sum_{k \in Z} x^{\pm}_j(k) z^{-k-1}
                        \quad (1 \leq j \leq n-1) . $$
   We define the action of $U_q \,$.
   $$ \gamma \longmapsto q \; \; \;,
      \; \;
      K_j \longmapsto q^{\partial_{\alpha_j}}
                        \quad (1 \leq j \leq n-1) , $$
   $$ X^{\pm}_j(z) \longmapsto
       {\rm exp}(\pm \sum^{\infty}_{k=1}
               \frac{a_j(-k)}{[k]} q^{\mp \frac{k}{2}} z^k)
       {\rm exp}(\mp \sum^{\infty}_{K=1}
               \frac{a_j(k)}{[k]} q^{\mp \frac{k}{2}} z^{-k})
       e^{\pm \alpha_j} z^{\pm \partial_{\alpha_j}} . $$
   We know the following theorem.
   \\
   {\bf Theorem 2.4}(\cite{fj})
   {\it
   By the above action,
   $W_i$ becomes
   the irreducible highest weight module
   with highest weight $\Lambda_i$,
   and
   $1 \otimes e^{\overline{\Lambda}_i}$
   is a highest weight vector of $W_i$.}
   \hspace{\fill} $\Box$
   \\
   From now on,
   we identify
   $W_i$
   and
   $1 \otimes e^{\overline{\Lambda}_i}$
   with
   $V(\Lambda_i)$ and $|\Lambda_i \rangle$
   respectively.


  \section{Construction of Vertex operators}

   In this section, we costruct
   the vertex operators on $W_i$
   explicitly.


   \subsection{Vertex operators}

    We review
    the definition and  some properties
    of the vertex operators.(\cite{fr},\cite{djo})
    Let V be a finite dimensional representation of $U'_q$.
    The affinization of $V$
    is the following $U_q$-module $V_z$.
    $$ V_z = V \otimes {\bf C}[z,z^{-1}] . $$
    We define
    the $U_q$-module structure on $V_z$
    as follows.
    $$ e_i. (v \otimes z^m) = e_i. v \otimes z^{m+\delta_{i0}}
       \quad , \quad
       f_i. (v \otimes z^m) = f_i. v \otimes z^{m-\delta_{i0}}  $$
    $$ t_i. (v \otimes z^m) = t_i. v \otimes z^m
       \quad , \quad
       q^d. (v \otimes z^m) = m v \otimes z^m.     $$
    \\
    {\bf Definition.}
    {\it
    The vertex operator is
    a $U_q$-homomorphism of the following form.

    Type I :
    $$ \tilde{\Phi}^{\mu V}_{\lambda}(z):
          V(\lambda) \longrightarrow V(\mu) \widehat{\otimes} V_z     $$

    Type II:
    $$ \tilde{\Phi}^{V \mu}_{\lambda}(z):
          V(\lambda) \longrightarrow V_z \widehat{\otimes} V(\mu)     $$}
    \hspace{\fill} $\Box$
    \\
    The symbol $\widehat{\otimes}$
    means
    $\widehat{W_1 \otimes W_2}$.
    From now  on, we omit it.
    We know the following theorem about
    the exisitence of the vertex operators.
    \\
    {\bf Theorem 3.1}(\cite{djo})
    \[ {\rm Hom}_{U_q}
         (V(\lambda),V(\mu) \otimes V_z) \]
    \[ \hspace{.8in}
       \cong
       \{ v \in V | {\rm \; the \; weight \; of \;} v
                         =\lambda - \mu {\rm \; mod \;} \delta \]
    \[ \hspace{1.5in}
                    {\rm and \; }
                          e_i^{\langle \mu , h_i \rangle +1}.v=0
                       {\rm \; for \;} i=0 , \cdots , n-1 \} , \]
    {\it
    where
    $\Phi \in {\rm Hom}_{U_q}(V(\lambda),V(\mu) \otimes V_z)$
    corresponds to $v$ via the relation
    $ \Phi | \lambda \rangle
          = | \mu \rangle \otimes v +$
    ( terms of positive powers in $z$ ).}
    \hspace{\fill} $\Box$
    \\
    We define
    the components of the vertex operators
    as follows.
    $$ \tilde{\Phi}^{\mu V}_{\lambda}(z) |u \rangle
           = \sum^{n-1}_{j=0}
             \tilde{\Phi}^{\mu V}_{\lambda}{}_j (z)
             |u \rangle \otimes v_j
         \qquad {\rm for} \quad |u \rangle \in V(\lambda) , $$
    where $ \{ v_j \} $ is a set of basis of V.
    For the type II, the components are also defined similarly.
    Using the components,
    we define similar vertex operators
    $$ \tilde{\Phi}^{\mu}_{\lambda V}(z):
          V(\lambda) \otimes V_z
           \longrightarrow V(\mu) \otimes {\bf C}[z,z^{-1}] $$
    by
    $$ \tilde{\Phi}^{\mu}_{\lambda V}(z) (|v \rangle \otimes v_i)
           =\tilde{\Phi}^{\mu V^*}_{\lambda}{}_j (z)|v \rangle
         \qquad {\rm for} \quad |v \rangle \in V(\lambda) . $$
    Here $x \in U_q$ acts on $V(\mu) \otimes {\bf C}[z,z^{-1}]$
    as $ x \otimes 1 $.

    Now, we specialize $V$ to the vector representation.
    $$ V = {\bf C}v_0 \oplus \cdots \oplus {\bf C}v_{n-1} $$
    The $U'_q$-module structure on $V$
    is the following.
    $$ e_i.v_j = \delta_{ij} v_{i-1}
       \quad , \quad
       f_i.v_j = \delta_{i-1j} v_i
       \quad , \quad
       t_i.v_j = q^{\delta_{ij+1}-\delta_{ij}} v_j . $$
    $(V^*)_z$ is denoted by $V_z^*$.
    The action of $U_q$ on $V_z^*$ is the following.
    $$ e_i.(v_j^* \otimes z^m)
          = - q^{-1} \delta_{i-1j} v_i^* \otimes z^{m+\delta_{i0}}
       \quad , \quad
       f_i.(v_j^* \otimes z^m)
          = - q \delta_{ij} v_{i-1}^* \otimes z^{m-\delta_{i0}} $$
    $$ t_i.v_j \otimes z^m=q^{-\delta_{ij+1}+\delta_{ij}} v_j \otimes z^m
       \quad , \quad q^d.v_j \otimes z^m=m v_j \otimes z^m.     $$
    In our case, by the above theorem, only
    $$ \tilde{\Phi}^{\Lambda_i V}_{\Lambda_{i+1}}(z)
       \; , \;
       \tilde{\Phi}^{\Lambda_{i+1} V^*}_{\Lambda_i}(z)
       \; , \;
       \tilde{\Phi}^{V \Lambda_i}_{\Lambda_{i+1}}(z)
       \; , \;
       \tilde{\Phi}^{V^* \Lambda_{i+1}}_{\Lambda_i}(z) $$
    are non-trivial.
    Furthermore, each of them
    is unique
    up to a scalar.
    Here, we take the following normalization.
    $$ \tilde{\Phi}^{\Lambda_i V}_{\Lambda_{i+1}}(z)
       | \Lambda_{i+1} \rangle
          =|\Lambda_i \rangle \otimes v_i
             + ( {\rm \; terms \; of
                      \; positive \; powers \; in} \; z \; ) $$
    $$ \tilde{\Phi}^{\Lambda_{i+1} V^*}_{\Lambda_i}(z)
       | \Lambda_i \rangle
          =|\Lambda_{i+1} \rangle \otimes v_i^*
             + ( {\rm \; terms \; of
                      \; positive \; powers \; in} \; z \; ) $$
    For the type II, we take a similar normalization.


   \subsection
         {Coproduct of $a_i(k),x^{\pm}_i(l)$ and
          action of $a_i(k),x^{\pm}_i(l)$ on $V_z$}

    The coproduct of Drinfeld generators
    is not known in full.
    But the ``main terms'' are calculated in \cite{cp}
    for $U_q(\widehat{sl}(2))$.
    The case of $U_q(\widehat{sl}(n))$
    is quite similar.
    \\
    {\bf Proposition 3.2.A}
    {\it
    For $k \geq 0$ , $l>0$ ,

    \[ \Delta (x^+_i(k))
          = x^+_i(k) \otimes \gamma^k + \gamma^{2k} K_i \otimes x^+_i(k) \]
    \[ \hspace{.75in}
            + \sum^{k-1}_{j=0} \gamma^{(k-j)/2} \psi_i(-k+j)
                               \otimes \gamma^{k-j} x^+_i(j)
              \qquad  {\it mod} \; UN_- \otimes UN_+^2       \]
    \[ \Delta (x^+_i(-l))
          = x^+_i(-l) \otimes \gamma^{-l} + K_i^{-1} \otimes x^+_i(-l) \]
    \[ \hspace{.75in}
            + \sum^{l-1}_{j=1} \gamma^{(l-j)/2} \varphi_i(-l+j)
                               \otimes \gamma^{-l+j} x^+_i(-j)
              \quad   {\it mod} \; UN_- \otimes UN_+^2       \]

    \[ \Delta (x^-_i(l))
          = x^-_i(l) \otimes K_i + \gamma^k \otimes x^-_i(l) \]
    \[ \hspace{.75in}
            + \sum^{k-1}_{j=1} \gamma^{k-j} x^-_i(j)
                               \otimes \gamma^{(j-k)/2} \psi_i(k-j)
              \qquad  {\it mod} \; UN_-^2 \otimes UN_+       \]
    \[ \Delta (x^-_i(-k))
          = x^-_i(-k) \otimes \gamma^{-2k} K_i^{-1}
                             + \gamma^{-k} \otimes x^-_i(-k) \]
    \[ \hspace{.7in}
            + \sum^{k-1}_{j=0} \gamma^{j-k} x^-_i(-j)
                               \otimes \gamma^{-(k+3j)/2} \varphi_i(j-k)
              \quad   {\it mod} \; UN_-^2 \otimes UN_+       \]

    \[ \Delta (a_i(l))
          = a_i(l) \otimes \gamma^{\frac{l}{2}}
            + \gamma^{\frac{3l}{2}} \otimes a_i(l)
              \qquad {\it mod} \; UN_- \otimes UN_+          \]
    \[ \Delta (a_i(-l))
          = a_i(-l) \otimes \gamma^{-\frac{3l}{2}}
            + \gamma^{-\frac{l}{2}} \otimes a_i(-l)
              \quad  {\it mod} \; UN_- \otimes UN_+          \]
    where, $UN_{\pm}$ , $UN_{\pm}^2$
            are the left ideals generated by
            $ \{ x_i^{\pm}(k) \} $, $ \{ x_i^{\pm}(k)x_j^{\pm}(l) \} $ .}

    \hspace{\fill} $\Box$
    \\
    {\bf Proposition 3.2.B}
    {\it
    The action of $ a_i(k),x^{\pm}_i(l) $ on $V_z$ is the following.
    $$ x^+_i(k) \longmapsto (q^i z)^k E_{ii-1} $$
    $$ x^-_i(k) \longmapsto (q^i z)^k E_{i-1i} $$
    $$ a_i(l)   \longmapsto
                \frac{[k]}{k}(q^i z)^k (q^{-k} E_{i-1i-1} - q^k E_{ii}) $$
    where
    $E_{ij}$ is the matrix unit of End$V$
    such that
    $ E_{ij}v_l=\delta_{jl}v_i $.}

    \hspace{\fill} $\Box$


   \subsection{Vertex operators of type I}

    First, we consider the vertex operator
    $\tilde{\Phi}^{\Lambda_i V}_{\Lambda_{i+1}}(z) :
       V(\Lambda_{i+1})
       \longrightarrow
       V(\Lambda_i) \otimes V_z$.
    We can determine the (n-1)-th component as follows.
    By prop.3.2,
    we get the following commutation relations.
    $$ [\tilde{\Phi}^V_{n-1}(z),X^+_j(w)]=0 $$
    $$ t_j \tilde{\Phi}^V_{n-1} t_j^{-1}
                 = q^{\delta_{j,n-1}} \tilde{\Phi}^V_{n-1}(z)  $$
    $$ [a_j(k),\tilde{\Phi}^V_{n-1}(z)]
        = \delta_{j,n-1} q^{\frac{2n+3}{2} k} \frac{[k]}{k} z^k
                                              \tilde{\Phi}^V_{n-1}(z)   $$
    $$ [a_j(-k),\tilde{\Phi}^V_{n-1}(z)]
        = \delta_{j,n-1} q^{-\frac{2n+1}{2} k} \frac{[k]}{k} z^{-k}
                                               \tilde{\Phi}^V_{n-1}(z)  $$
    \hspace{\fill} for $ 1 \leq j \leq n-1 $.
    \\
    The above conditions determine
    the form of
    $\tilde{\Phi}^{\Lambda_i V}_{\Lambda_{i+1} n-1}(z)$
    completely
    under the normalization conditions in \S 3.1.
    The other components
    are determined
    by one of the intertwining conditions.
    $$ \tilde{\Phi}^{\Lambda_i V}_{\Lambda_{i+1}}{}_{j-1}(z)
       =[ \; \tilde{\Phi}^{\Lambda_i V}_{\Lambda_{i+1}}{}_{j}(z)
          \; , f_j \; ]_q . $$
    Hence,
    the other components are
    represented by the integral of the currents.

    For the vertex operators
    $ \tilde{\Phi}^{\Lambda_{i+1} V^*}_{\Lambda_i}(z) :
      V(\Lambda_i)
      \longrightarrow
      V(\Lambda_{i+1}) \otimes {V_z}^* $,
    we have
    the similar commutaion relations
    this time
    for the 0-th component.
    We summarize the results.
    \\
    {\bf Theorem 3.3}
    \\
    1) $ \; \displaystyle{
      \tilde{\Phi}^{\Lambda_i V}_{\Lambda_{i+1} n-1}(z)
        ={\rm exp}(\sum^{\infty}_{k=1}
               a^*_{n-1}(-k) q^{\frac{2n+3}{2} k} z^k)
         {\rm exp}(\sum^{\infty}_{k=1}
               a^*_{n-1}(k) q^{-\frac{2n+1}{2} k} z^{-k})
                         } $
    \[   \hspace{1.75in} \times
         e^{\overline{\Lambda}_{n-1}}
         (q^{n+1} z)^{\partial_{\overline{\Lambda}_{n-1}}+\frac{n-i-1}{n}} \]
    \[   \hspace{1.75in} \times
         (-1)^{(\partial_{\overline{\Lambda}_1}-\frac{n-i-1}{n})(n-1)}
         (-1)^{\frac{1}{2} (n-i)(n-i-1)}                        \]
    $    \hspace*{\fill} (i=0,\cdots,n-1) $
    $$ \tilde{\Phi}^{\Lambda_i V}_{\Lambda_{i+1} j-1}(z)
       =[ \; \tilde{\Phi}^{\Lambda_i V}_{\Lambda_{i+1} j}(z)
          \; , f_j \; ]_q
       \quad (j=1 , \cdots , n-1) . $$
    2) $ \; \displaystyle{
      \tilde{\Phi}^{\Lambda_{i+1}}_{\Lambda_i V 0}(z)
           ={\rm exp}(\sum^\infty_{k=1} a^*_1 (-k) q^{\frac{3}{2} k} z^k)
            {\rm exp}(\sum^\infty_{k=1} a^*_1 (k) q^{-\frac{1}{2} k} z^{-k})
                         } $
    \[ \hspace{1.75in} \times
            e^{\overline{\Lambda}_1}
            ((-1)^{n-1}qz)^{\partial_{\overline{\Lambda}_1}+\frac{i}{n}}
            q^i (-1)^{in+\frac{1}{2} i(i+1)}       \]
    $    \hspace*{\fill} (i=0,\cdots,n-1) $
    $$ \tilde{\Phi}^{\Lambda_{i+1} V^*}_{\Lambda_i j}(z)
           =[ \; f_j \; , \;
              \tilde{\Phi}^{\Lambda_{i+1} V^*}_{\Lambda_i j-1}(z) \; ]_{q^{-1}}
       \quad (j=1 , \cdots , n-1) .     $$
    {\it
    where
    $ \displaystyle{
        a^*_{n-1}(k)=\sum^{n-1}_{l=1} \frac{-[lk]}{[k][nk]} a_l(k) , \;
        a^*_1 (k)=\sum^{n-1}_{l=1}
                                 \frac{-[(n-l)k]}{[k][nk]} a_l (k) .} $
    \\
    The coffiecients of
    $a^*_{n-1}(k)$
    and
    $a^*_1 (k)$
    are determined by the conditions
    $$ [a_i(k),a^*_{n-1}(-k)]=\delta_{i,n-1} \frac{[k]}{k} \; , \;
       [a_i(k),a_1^*(-k)] = \delta_{i1} \frac{[k]}{k} . $$}
    \hspace*{\fill} $\Box$


   \subsection{Vertex operators of type II}

    We can also apply
    the same method for the vertex operators of type II.
    \\
    {\bf Theorem 3.4}
    \\
    1) $ \; \displaystyle{
         \tilde{\Phi}^{V \Lambda_i}_{\Lambda_{i+1} 0}(z)
           ={\rm exp}
             (-\sum^\infty_{k=1} a^*_1 (-k) q^{\frac{1}{2} k} z^k)
            {\rm exp}
             (-\sum^\infty_{k=1} a^*_1 (k) q^{-\frac{3}{2} k} z^{-k})
                         } $
    \[ \hspace{1.4in} \times
            e^{-\overline{\Lambda}_1}
            ((-1)^{n-1}qz)^{-\partial_{\overline{\Lambda_1}}+\frac{n-i-1}{n}}
            q^{-i} (-1)^{in+\frac{1}{2} i(i+1)}       \]
    \hspace{\fill} $(i=0,\cdots,n-1)$
    $$ \tilde{\Phi}^{V \Lambda_i}_{\Lambda_{i+1} j}(z)
       =[ \; \tilde{\Phi}^{V \Lambda_i}_{\Lambda_{i+1} j-1}(z)
          \; , e_j \; ]_q
       \quad (j=1 , \cdots , n-1) . $$
    2) $ \; \displaystyle{
       \tilde{\Phi}^{V^* \Lambda_{i+1}}_{\Lambda_i n-1}(z)
        ={\rm exp}
          (-\sum^{\infty}_{k=1} a^*_{n-1}(-k) q^{\frac{2n+1}{2} k} z^k)
         {\rm exp}
          (-\sum^{\infty}_{k=1} a^*_{n-1}(k) q^{-\frac{2n+3}{2} k} z^{-k})
                         }$
    \[   \hspace{1.4in} \times
         e^{-\overline{\Lambda}_{n-1}}
         (q^{n+1} z)^{-\partial_{\overline{\Lambda}_{n-1}}+\frac{i}{n}} \]
    \[   \hspace{1.4in} \times
         (-1)^{(\partial_{\overline{\Lambda}_1}-\frac{n-i}{n})(n-1)}
         (-1)^{\frac{1}{2} (n-i)(n-i-1)}                        \]
    \hspace{\fill} $(i=0,\cdots,n-1)$
    $$ \tilde{\Phi}^{V^* \Lambda_{i+1}}_{\Lambda_{i+1} j-1}(z)
       =[ \; e_j \; ,
          \; \tilde{\Phi}^{V^* \Lambda_i}_{\Lambda_{i+1} j}(z)
        \; ]_{q^{-1}}
       \quad (j=1 , \cdots , n-1) . $$
    \hspace*{\fill} $\Box$


   \subsection{Commutation relations of the vertex operators}

   In \cite{do},
   by solving the $q$-KZ equations
   the authors get
   the commutation relations
   of the vertex operators of type I
   related to $U_q(\widehat{sl}(n))$.
   These formulas can be derived directry
   by using our explicit formulas for vertex operators
   First, we write down
   the matrix coffiecient of
   $  \overline{R}_{V^* V}(z_1/z_2)
           \in {\rm End}_{\bf C} V^*_{z_1} \otimes V_{z_2} $.
   $$ \overline{R}_{V^* V}(z)(v^*_i \otimes v_j)
        =v_i^* \otimes v_j \; (i \neq j) , \;
      \overline{R}_{V^* V}(z)(v^*_i \otimes v_i)
        =\sum^{n-1}_{j=0} a_{ij} v_j^* \otimes v_j , $$
   $$ {\rm where} \quad
      a_{ij}=
        \left \{
        \begin{array}{ll}
         {\displaystyle \frac{(q-q^{-1})z}{1-z}}
       & (i>j) \\
         {\displaystyle \frac{q-q^{-1}z}  {1-z}}
       & (i=j) \\
         {\displaystyle \frac{q-q^{-1}}   {1-z}}
       & (i<j) \quad .
        \end{array}
        \right. $$
   \\
   {\bf Proposition 3.5}
   \[ 1) \hspace{.75in}
           \tilde{\Phi}^{\Lambda_{i+1}}_{\Lambda_i V}(z)
           \tilde{\Phi}^{\Lambda_i V}_{\Lambda_{i+1}}(z)
           =\frac{(q^{2n};q^{2n})_\infty}{(q^2;q^{2n})_\infty}
            id_{V(\Lambda_{i+1})}           \]
   \[ 2) \hspace{.75in}
           \tilde{\Phi}^{\Lambda_i V}_{\Lambda_{i+1}}(z)
           \tilde{\Phi}^{\Lambda_{i+1}}_{\Lambda_i V}(z)
           =\frac{(q^{2n};q^{2n})_\infty}{(q^2;q^{2n})_\infty}
            id_{V(\Lambda_i) \otimes V} \]
   \[ 3) \hspace{.5in}
           \tilde{\Phi}^{\Lambda_i V}_{\Lambda_{i+1}}(z_2)
           \tilde{\Phi}^{\Lambda_{i+1} V^*}_{\Lambda_i}(z_1) \]
   \[    \hspace{1in}
           = -q(\frac{z_1}{z_2})^{-\delta_{i0}}
              r(\frac{z_1}{z_2}) P \overline{R}_{V^* V}(\frac{z_1}{z_2})
              \tilde{\Phi}^{\Lambda_i V^*}_{\Lambda_{i-1}}(z_1)
              \tilde{\Phi}^{\Lambda_{i-1} V}_{\Lambda_i}(z_2)   \]
        $$ {\it where} \quad
             r(z)
           = \frac{(z;q^{2n})_\infty (q^{2n+2} z^{-1};q^{2n})_\infty}
                  {(q^2 z;q^{2n})_\infty (q^{2n} z^{-1};q^{2n})_\infty}
             \; , \;
           P v^*_i \otimes v_j = v_j \otimes v^*_i . $$
   $proof)$
          Formulas 1), 2) follow from simple calculations.
          We know the uniqueness of the vertex operator
          $ V(\Lambda_i) \longrightarrow
                V(\Lambda_i) \otimes V^*_{z_1} \otimes V_{z_2} $.
          (For the details see \cite{fr}\cite{djo}.)
          So, the left and the right hand sides of 3)
          coincide up to a scalar factor.
          By comparing the $v_n \otimes v_1^*$ component of both sides,
          we get the above equation.
          \hspace{\fill} $\Box$


  \section{Vertex model}

   In this section,
   we give a mathematical definition of the model
   treated in this paper.
   (\cite{dfjmn},\cite{iijmnt},\cite{do})


   \subsection{Space of states}

    We know
    the integrable generalization of
    the $XXZ$-model related to
    any quantum affine algebra
    $U_q(\hat g)$.
    Let $V_z$ be
    a finite dimensional representation of $U'_q(\hat g)$
    with a spectral parameter $z$
    and
    $R(z_1/z_2) \in$End$(V_{z_1} \otimes V_{z_2})$
    be the $R$-matrix for $U'_q(\hat g)$.
    We define the model on
    the infinite lattice
    $ \cdots \otimes V \otimes V \otimes V \otimes \cdots $.
    Let $h$ be the operator on $V \otimes V$
    such that
    $$ PR(z_1/z_2)=(1+uh+ \cdots ) \times {\rm const.}
                                \quad (u \rightarrow 0), $$
    $$ P \; : \; {\rm the \; transposition}, \quad e^u=z_1/z_2 $$
    We define the Haniltonian ${\cal H}$ as follows.
    $$ {\cal H}=\sum_{k \in {\bf Z}} h_{l+1 l} , $$
    where $ h_{l+1 l} $ is
    $ \cdots \otimes 1 \otimes 1
           \otimes h \otimes 1 \otimes 1 \otimes \cdots $
    acting the $l$-th component and $l+1$-th component.
    We can check immediately
    $$ [ \, U'_q(\hat g) \, , \, {\cal H} \, ]=0 . $$
    When
    $ g =sl(2) $
    and $V_z$ is two-dimensional $U'_q(\widehat{sl}(2))$-module,
    ${\cal H}$ becomes $H_{XXZ}$.

    From now on,
    we specialize
    $g$ to $sl(n)$
    and
    $V_z$ to the vector representation of $U'_q$
    in \S 3.1.
    Later,
    when we solve
    the difference equations for the one-point functions,
    we find it convinient
    to pass to an equivalent representation
    $V^{pr}_{\zeta}$ defined by
    $$ V={\bf C} u_1 \oplus \cdots \oplus {\bf C} u_{n-1} , $$
    $$ e_i.u_j = \delta_{ij} u_{i-1} \zeta , \;
       f_i.u_j = \delta_{i-1j} u_i \zeta^{-1} , \;
       t_i.u_j = q^{\delta_{i-1j} - \delta_{ij}} u_j . $$
    The equivalence is given by
    $$ V_z \longrightarrow V^{pr}_{\zeta}
       , \;
       v_i \longmapsto u_i \zeta^{-i}
       , \; z=\zeta^n . $$
    We will refer to
    $V_z$ and $V^{pr}_{\zeta}$
    as the homogineneous picture
    and the principal picture, respectively.
    As explained in the introduction, we take
    $$ {\rm End}_{\bf C}
      ( \bigoplus_{i=0}^{n-1} V({\Lambda}_i) )
      \cong
      \bigoplus_{i,j} V({\lambda}_i){\otimes}V({\lambda}_j)^* $$
    as the space of states ${\cal F}$.
    ${\cal F}$
    is understood naively
    as the subspace of the infinite tensor product
    $ \cdots
      {\otimes}V{\otimes}V{\otimes}V{\otimes}
      \cdots $.
    We give
    the left and right action of $U$ on ${\cal F}$
    as follows.
    $$ x.f=\sum x_{(1)} \circ f \circ S(x_{(2)})      $$
    $$ f.x=\sum S^{-1}(x_{(2)}) \circ f \circ x_{(1)} $$
    $$ {\rm where} \qquad
         f \in {\cal F} \; , \; x \in U \; , \;
         \Delta (x)=\sum x_{(1)} \otimes x_{(2)} \; . $$
    The space ${\cal F}$ regarded as the right module
    is denoted by ${\cal F}^r$.
    Let
    $$ {\cal F}_{ij}
         ={\rm Hom}(V({\lambda}_j),V({\lambda}_i))
                \cong V({\lambda}_i){\otimes}V({\lambda}_j)^* . $$
    ${\cal F}_{ii}$ has
    the unique canonical element $id_{V(\Lambda_i)}$.
    We call it the vacuum and denote it by
    $|vac \rangle _i \in {\cal F}_{ii} \; ,
    \; _i \langle vac| \in {\cal F}_{ii}^r$.
    There is a natural inner product between
    ${\cal F}_{ij}^r$ and ${\cal F}_{ji}$ as follows.
    $$ \langle f|g \rangle
        = \frac{{\rm tr}_{V(\Lambda_i)}(q^{-2 \rho}fg)}
               {{\rm tr}_{V(\Lambda_i)}(q^{-2 \rho})}
       \qquad {\rm for} \;
       f \in {\cal F}_{ij}^r \; , \; g \in {\cal F}_{ji} \; , $$
    $$ {\rm where} \quad
       \rho
        = \Lambda_0 + \Lambda_1 + \cdots + \Lambda_{n-1} \; . $$
    It is invariant under the action of $U_q$ :
    $ \langle fx|g \rangle
                = \langle f|xg \rangle$ for $^{\forall} x \in U $.


   \subsection{Local structure and local operators}

    We use the vertex operator
    $$ \tilde{\Phi}^{\Lambda_{i-1} V}_{\Lambda_i}(z):
       V(\Lambda_i) \longrightarrow V(\Lambda_{i-1}) \otimes V_z $$
    to incorporate
    the local structure into
    ${\cal F}$.
    \\
    Setting $z=1$, we obtain
    the $U'_q$-homomorphism
    $$ \tilde{\Phi}^{\Lambda_{i-1} V}_{\Lambda_i}:
       V(\Lambda_i) \longrightarrow \widehat{V(\Lambda_{i-1})} \otimes V . $$
    Let
    $$ \tilde{\Phi}^{(m)}_{\Lambda_i}
            :=\tilde{\Phi}^{\Lambda_{i-m} V}_{\Lambda_{i-m+1}}
              \cdots
              \tilde{\Phi}^{\Lambda_{i-2} V}_{\Lambda_{i-1}}
              \tilde{\Phi}^{\Lambda_{i-1} V}_{\Lambda_i} .     $$
    $ \tilde{\Phi}^{(m)}_{\Lambda_i} $
    converges and
    gives the following isomorphism.
    $$ {\cal F}_{ij}=V(\Lambda_i) \otimes V(\Lambda_j)^*
           \cong V(\Lambda_{i-m}) \otimes
        \underbrace{V \otimes \cdot \cdot \cdot \otimes V}_{\rm m-times}
        \otimes V(\Lambda_j)^* . $$
    By this isomorphism,
    the local structure
    is inserted into ${\cal F}$.
    Next, we define the local operators.
    For $L \in $End$ V^{\otimes m}$,
    let
    $$ {\cal L}_{(i)}:={(\tilde{\Phi}^{(m)}_{\Lambda_i})}^{-1}
                       (id_{V(\Lambda_{i-m})} \otimes L)
                       (\tilde{\Phi}^{(m)}_{\Lambda_i}) .     $$
    By prop.3.6, we know
    $$ {(\tilde{\Phi}^{(m)}_{\Lambda_i})}^{-1}
           =\left(
             \frac{( q^2  ;q^{2n})_{\infty}}
                  {(q^{2n};q^{2n})_{\infty}}
             \right) ^m
              \tilde{\Phi}^{\Lambda_i}_{\Lambda_{i-1} V}
              \tilde{\Phi}^{\Lambda_{i-1}}_{\Lambda_{i-2} V}
              \cdots
              \tilde{\Phi}^{\Lambda_{i-m+1}}_{\Lambda_{i-m} V} , $$
    where
    $\tilde{\Phi}^{\Lambda_i}_{\Lambda_{i-1} V}
      =\tilde{\Phi}^{\Lambda_i}_{\Lambda_{i-1} V}(1)$.
    The action of $L$ on ${\cal F}_{ij}$ is defined as follows.
    $$ L.f:={\cal L}_{(i)} \circ f . $$
    We denote the correlator $_i \langle vac|L|vac \rangle_i $
    by $ \langle L \rangle^{(i)}$.


  \section{Staggered polarization}

   The aim of this section is to give
   $ \langle E_{m'm} \rangle^{(i)} $
   explicitly.


   \subsection{Integral representations}

    In \cite{jmmn},
    the authors construct
    an integral representation
    of correlators of the $XXZ$-model
    by using the trace formula
    explained in \cite{cs} Appendix C.
    We can apply
    the same method to
    $ \langle E_{m'm} \rangle^{(i)} $.
    \\
    Put
    $$ P^m_{m'}(z_1,z_2|x,y|i)
       := \frac{(q^2;q^{2n})_\infty}
              {(q^{2n};q^{2n})_\infty}
         \frac{{\rm tr}_{V(\Lambda_i)}
                     (x^{-d} y^{2 \overline{\rho}}
                      \tilde{\Phi}^{\Lambda_i}
                                  _{\Lambda_{i-1} V m'}(z_1)
                      \tilde{\Phi}^{\Lambda_{i-1} V}
                                  _{\Lambda_i m}(z_2))}
              {{\rm tr}_{V(\Lambda_i)}
                    (x^{-d} y^{2 \overline{\rho}})} , $$
    then \,
    $ \langle E_{m' m} \rangle^{(i)}
       = P^m_{m'}(z,z|q^{2n},q^{-1}|i) $.
    \\
    Let
    $$ h(z)=(z;x)_\infty (q^2 z^{-1};x)_\infty , $$
    $$ \varphi (z ; x)
            = \prod^\infty_{k=1}
                \frac{(z x^k                  ;q^{2n})_\infty
                      (q^{2n} z^{-1} x^{k-1}  ;q^{2n})_\infty}
                     {(q^2 z x^k              ;q^{2n})_\infty
                      (q^{2n+2} z^{-1} x^{k-1};q^{2n})_\infty} , $$
    $$ \theta_i(z_1,\cdots ,z_{n-1})
          =y^{(2\overline{\rho}|\overline{\Lambda}_i)}
            \sum_{\alpha \in \overline{Q}}
                x^{\frac{(\alpha|\alpha)}{2}
                         +(\alpha|\overline{\Lambda}_i)}
                z_1^{(\overline{\Lambda}_1|\alpha)} \cdots
                z_{n-1}^{(\overline{\Lambda}_{n-1}|\alpha)} . $$
    We get the following.
    \[ P^m_{m'}(z_1,z_2|x,y|i) \]
    \[ \hspace{.4in}
         =c \times
          \frac{\delta_{m'm}
                (q^2;q^{2n})_\infty (q^2;x)_\infty^{n-1}
                \varphi (z;x) }
               {(q^{2n};q^{2n})_\infty
                tr_{V(\Lambda_i)}(x^{-d} y^{2 \overline{\rho}})}
          \left(
           \prod^\infty_{k=1}
           \frac{(q^2 x^k;q^{2n})_\infty}
                {(q^{2n} x^k;q^{2n})_\infty}
          \right) ^2 \]
    \[ \hspace{.45in}
       \times
       \oint_{q^2<|w_l|<1 (l \neq m)}
         \frac{d \xi_1 \cdots d \xi_{n-1}}
              {(2 \pi \sqrt{-1})^{n-1}
                  \xi_1 \cdots \xi_{n-1}}
         (\frac{q}{\xi_i})^{1-\delta_{i0}}         \]
    \[ \hspace{.8in} \times
         \frac{(1-zw_m) w_{m+1} \cdots w_{n-1}}
              {h(w_0)
               \cdots
               h(w_{m-1})h(zw_m)h(w_{m+1})
               \cdots
               h(w_{n-1})} \]
    \[ \hspace{.8in} \times
         \theta_i (\eta_1 , \cdots , \eta_{m-1} ,
                   \eta_m z^{-1} , \eta_{m+1} z ,
                   \eta_{m+2} , \cdots , \eta_{n-1})  \]
    \[ {\rm where} \quad
       w_0 = \frac{q^2}{\xi_1} , \;
       w_1 = \frac{q \xi_1}{\xi_2} , \;
       \cdots , \;
       w_{n-2} = \frac{q \xi_{n-2}}{\xi_{n-1}} , \;
       w_{n-1} = \frac{\xi_{n-1}}{q^n},          \]
    \[ \hspace{.5in}
       z= \frac{z_1}{z_2} , \;
       \eta_j  = \frac{w_{j-1}}{w_j} y^2 , \;
       c=\left\{
            \begin{array}{ll}
             q^i        & ( m  <   i ) \\
             q^i z^{-1} & ( m \geq i ) .
            \end{array}
         \right. \]
    By this expression, we can verify that
    $$  {\varphi (z ; x)}^{-1}
        {\rm tr}_{V(\Lambda_i)}(q^{-2 \rho}
               \tilde{\Phi}^{\Lambda_i V^*}_{\Lambda_{i-1}}(z_1)
               \tilde{\Phi}^{\Lambda_{i-1} V}_{\Lambda_i}(z_2)) $$
    is a function of $z(=z_1/z_2)$
    and regular in $q^{-2n}<|z|<q^{2n}$.


   \subsection{Staggered polarization}

    In this subsection we derive
    the difference equations
    for one point functions
    and solve them.
    These equations can be solved easily
    up to a pseudo-constant factor
    and
    we can determin the factor
    by the analyticity
    gained from the integral representations.
    Following \cite{iijmnt},
    we explain
    how to derive
    the difference equations
    in our context.

    Let
    $$ \tilde{F}^{(i)} (\frac{z_1}{z_2})
        :={\rm tr}_{V(\Lambda_i)}(q^{-2 \rho}
               \tilde{\Phi}^{\Lambda_i V^*}_{\Lambda_{i-1}}(z_1)
               \tilde{\Phi}^{\Lambda_{i-1} V}_{\Lambda_i}(z_2))  $$
    then, we get the equation
    \[  \tilde{F}^{(i)} (\frac{z_1}{z_2 q^{2n}})
         ={\rm tr}_{V(\Lambda_i)}(q^{-2 \rho}
               \tilde{\Phi}^{\Lambda_i V^*}_{\Lambda_{i-1}}(z_1)
               \tilde{\Phi}^{\Lambda_{i-1} V^*}_{\Lambda_i}(z_2 q^{2n})) \]
    \[ \hspace{.745in}
         = P{\rm tr}_{V(\Lambda_{i-1})}
              (\tilde{\Phi}^{\Lambda_{i-1} V}_{\Lambda_i}(z_2 q^{2n})
               q^{-2 \rho}
               \tilde{\Phi}^{\Lambda_i V^*}_{\Lambda_{i-1}}(z_1))        \]
    \[ \hspace{.745in}
         = (q^{-2 \overline{\rho}} \otimes 1)
            P{\rm tr}_{\Lambda_{i-1}}(q^{-2 \rho}
               \tilde{\Phi}^{\Lambda_{i-1} V}_{\Lambda_i}(z_2)
               \tilde{\Phi}^{\Lambda_i V^*}_{\Lambda_{i-1}}(z_1))        \]
    \[ \hspace{.745in}
         = -q(\frac{z_1}{z_2})^{-\delta_{i1}} r(\frac{z_1}{z_2})
           (1 \otimes q^{-2 \overline{\rho}})
           \overline{R}_{V^* V}(\frac{z_1}{z_2})                         \]
    \[ \hspace{1.2in}
           \times
           {\rm tr}_{V(\Lambda_{i-1})}(q^{-2 \rho}
               \tilde{\Phi}^{\Lambda_{i-1} V^*}_{\Lambda_{i-2}}(z_1)
               \tilde{\Phi}^{\Lambda_{i-2} V}_{\Lambda_{i-1}}(z_2))      \]
    or,
    $$ \tilde{F}^{(i)}(z q^{-2n})
         = - qz^{-\delta_{i1}}
                (1 \otimes q^{-2 \overline{\rho}})
                 r(z) \overline{R}_{V^* V}(z) \tilde{F}^{(i-1)}(z) \; .  $$
    We show this equation reduces to scaler equations.
    Let
    $$ \overline{F}^{(i)}(z)
                 :=\varphi(z ; q^{2n})^{-1} \tilde{F}^{(i)}(z) , $$
    then
    $$ \overline{F}^{(i)}(z q^{-2n})
         = - qz^{-\delta_{i1}}
                (1 \otimes q^{-2 \overline{\rho}})
                 \overline{R}_{V^* V}(z) \overline{F}^{(i-1)}(z) \; . $$
    Let
    $ z=\zeta^n $
    and
    $ \omega $
    be an n-th primitive root of 1.
    We put
    $$ \sum^{n-1}_{m=0}
          G^{(j)}_m(\zeta) v^*_m \otimes v_m
           :=\sum^n_{i=1} q^{(n-i)i}
               \omega^{ij} \zeta^{n-i} \overline{F}^{(i)}(\zeta^n) . $$
    \\
    Let further
    $$ G^{(j,k)}(\zeta)
          := \zeta \sum^{n-1}_{m=0} \omega^{km}
                      \zeta^{m-n} G^{(j)}_m (\zeta).   $$
    Then, we find
    $$ \frac{G^{(j,k)}(\zeta q^{-2})}
            {1-\omega^k \zeta q^{-2}}
      = - q^2 \omega^j \zeta^{-1}
          \frac{G^{(j,k)}(\zeta)}{1 - \omega^k \zeta} \; . $$
    Let
    $$ \Theta_p(z)
            =(p;p)_\infty (z;p)_\infty (z^{-1} p;p)_\infty \; . $$
    The reduced equation determines
    $G^{(j,k)}(\zeta)$
    as
    $$ G^{(j,k)}(\zeta)
           =c_{jk}(\zeta)
               \frac{1-\omega^k \zeta}
                    {\Theta_{q^2}(\omega^{-j} \zeta)}
       \qquad; (*) \; $$
    where $c_{jk}(\zeta)$ is a pseudo-constant
    ( i.e. $c_{jk}(\zeta q^{-2})=c_{jk}(\zeta)$ ).
    Let us show that
    $c_{jk}(\zeta)$ is independent of $zeta$.
    As $ \overline{F}(\zeta^n) $ is regular in $ q^{-2}<|\zeta|<q^2 $,
    so is $ G^{(j,k)}(\zeta) $.
    So, when we set
    $ \zeta=e^{2 \pi \sqrt{-1} u} $
    and
    $ q=e^{\pi \sqrt{-1} \tau} $,
    $$ c_{jk}(\zeta)=G^{(j,k)}(\zeta)
                      \frac{\Theta_{q^2}(\omega^{-j} \zeta)}
                           {1-\omega^k \zeta}              $$
    has at most a simple pole
    in the fundamental rigion $[0,1] \times [0,\tau]$
    in the $u$-plane.
    Hence, $ c_{jk}(\zeta) $ is
    an absolute-constant $c_{jk}$.
    Moreover, it can be determined
    by calculating the residue at
    $ \zeta=q^{-2} \omega^j $
    of the both sides of the above equation $(*)$.
    The result is the following.
    $$ G^{(j,k)}(\zeta)
          =\left \{
             \begin{array}{ll}
            {\displaystyle
                     n \omega^j C
                     \frac{1-\omega^{-j} \zeta}
                          {\Theta_{q^2}(\omega^{-j} \zeta)}}
                          & (j+k \equiv 0 \;{\rm mod} \; n) \\
                    0     & ({\rm otherwise})
             \end{array}
           \right. $$
    \hspace{1in} where
    $ \displaystyle{
       C=\frac{(q^2;q^2)_\infty^3 (q^{2n};q^{2n})_\infty}
              {\varphi(1 ; q^{2n}) (q^2;q^{2n})_\infty}
         {\rm tr}_{V(\Lambda_0)}(q^{-2 \rho})} .      $
    \\
    We get the following theorem.
    \\
    {\bf Theorem 5.2}
    {\it
    Let $\omega$ be an n-th primitive root of 1
    and
    $E_{ij}$ be the matrix unit,
    then}
    $$ \sum^{n-1}_{m=0}
       \omega^{km} \langle E_{mm} \rangle^{(i)}
            =\frac{\omega^{(i-1)k}(q^2;q^2)_\infty^2}
               {(q^2 \omega^k;q^2)_\infty
                (q^2 \omega^{-k};q^2)_\infty} \; \; . $$
    \hspace{\fill} $\Box$
    \\
    $Acknowledgement.$
    I wish to thank
    M.Jimbo, M.Kashiwara and T.Miwa
    for discussions and suggestions.



\end{document}